\documentstyle[multicol,aps,prl]{revtex}

\begin{document}
\pagestyle{empty}

\newcommand{\bc}{\begin{center}}
\newcommand{\ec}{\end{center}}
\newcommand{\be}{\begin{equation}}
\newcommand{\ee}{\end{equation}}
\newcommand{\beqn}{\begin{eqnarray}}
\newcommand{\eeqn}{\end{eqnarray}}
\newcommand{\bphi}{\bbox{\phi}}

\begin{multicols}{2}
\narrowtext
\parskip=0cm

\noindent
{\bf Emig and Nattermann Reply:} The Comment of Hazareesing and
Bouchaud \cite{HB} on our recent Letter 'Roughening Transition of
Interfaces in Disordered Systems' \cite{Emig1} addresses three
points:

(i) They obtain an order $\epsilon=4-D$ correction to our result
$\nu^{-1}=2\sqrt{\epsilon}$ for the correlation length exponent $\nu$.

(ii) They consider a {\em generalization of our model} to an $N$ dimensional
displacement field $\bphi=(\phi_1,\ldots,\phi_N)$ and find for that
model a first-order transition if $N>4$.

(iii) They criticize the generalization of our results to systems with
non-local elastic interaction by arguing that there is {\it no}
renormalization of the non-local elastic term and hence no fixed point
in lowest order $\epsilon$, which again may result in a first order
transition. 

Our Reply is as follows:

(i) The corrections of order $\epsilon$ to $\nu^{-1}$ are not only
determined by one-loop contributions -- as considered in Refs.
\CITE{HB,Emig1} -- but also by higher order terms which may lead also
to $\epsilon^{3/2}$ contributions to the disorder correlator
\cite{Balents}. The calculations presented in Refs. \CITE{HB,Emig1}
are only able to give the value of $\nu^{-1}$ to order
$\sqrt{\epsilon}$ correctly.

(ii) The generalized Hamiltonian of HB assumes a periodic potential $v
\cos(2\pi(\phi_1+\ldots+\phi_N)/\sqrt{N})$ which has the property that
even for $v \to \infty$ the manifold is rough in the $N-1$ dimensional
subspace $\sum_i \phi_i = n\sqrt{N}$ where $n$ is integer. Lowering
the ratio between $v$ and the disorder strength, eventually a
transition to a phase will occur in which the confinement to this
subspace is lost. However, note that in the latter phase the roughness
exponent is {\it smaller} than in the confined phase for larger $v$.
Even if this transition would be first order for $N>4$, as HB claim,
its physical significance is not completely evident.

In a separate article, we have considered a {\em different} generalization
to a $N>1$ component model \cite{Emig2} by assuming a periodic
potential of the form $v(\cos\phi_1+\ldots+\cos\phi_N)$, which is free
of the physical deficiencies of the model proposed by HB. Using a
functional renormalization group approach, our model shows a continous
transition for all finite $N$ and the result for the exponent $\nu$
given in our Letter remains valid to order $\sqrt{\epsilon}$
independent of $N$.

(iii) In the case of non-local elasticity the upper and lower critical
dimension for the appearance of a roughening transition decrease from
$D_u=4$, $D_l=2$ to $D_u=3$, $D_l=5/3$, respectively, as stated in
\cite{Emig1}. However, HB are correct in their statement that there is
indeed no renormalization of the non-local elastic term as we assumed
erroneously. This may result in a first order transition as found in
the self-consistent calculation by M\"ussel \cite{Muessel}, who
neglected a renormalization of the stiffness constant from the very
beginning.

We thank J.-P. Bouchaud for useful discussions.

\bigskip
\noindent
Thorsten Emig and Thomas Nattermann
\\
\\
{\small
Institut f\"ur Theoretische Physik\\ 
Universit\"at zu K\"oln\\
Z\"ulpicher Str. 77\\
\\
D-50937 K\"oln, Germany

}

\bigskip
\noindent
Date: 23. October 1998


\end{multicols}

\end{document}